# A SEARCH FOR THERMOSYNTHESIS: STARVATION SURVIVAL IN THERMALLY CYCLED BACTERIA

## Anthonie W.J. Muller

Department of Geology, Washington State University,

Pullman WA 99164-2812, USA

Tel: + 509-335-1501; fax: + 509-335-7816;

E-mail address: awjmuller@wsu.edu

**Abstract**

In a pioneering study experimental evidence was sought of thermosynthesis, a theoretical biological mechanism for free energy gain from thermal cycling that has been invoked as energy source for the origin of life. A PCR machine applied thermal cycling to the K12 strain of *Escherichia coli*. The viability of this organism during starvation was determined at cyclic and at constant temperature. The found increase in the viability counts during the first days of starvation is consistent with thermosynthesis. The scattering in the results is however large. Further research is needed to proof that the increase is indeed due to a thermosynthesis process.

Thermosynthesis, Origin of Life, Starvation, Quorum Sensing

## INTRODUCTION

From December 2004 to January 2006 the author worked as visiting assistant professor in Prof. Schulze-Makuch's lab at the Geology Department of the Washington State University in Pullman. The goal was to obtain experimental evidence for the theoretical notion of thermosynthesis, biological free energy gain from thermal cycling.[1] This document reports on this endeavor.

The simplest form of thermosynthesis uses a single protein that resembles the $F_1$ subunit of today's ATP synthase enzyme. The simplicity has been applied in models for (i) the emergence of the genetic machinery in the RNA World[2] and (ii) the emergence of the chemiosmotic energy conversion machinery.[3] In thermosynthesis a free energy-yielding enzymatic cycle is synchronized with a thermal cycle. Since in the natural environments thermal cycling times, such as during convection (~ 100 s), are in general much larger than turnover times of enzymes, for ATP synthase normally ~ 0.01 s, the power of thermosynthesis can only be small compared to that of regular processes such as respiration. This low power makes thermosynthesis difficult to detect.

Nevertheless several direct experimental approaches for the detection of thermosynthesis have been proposed based on sensitive detection methods of small molecules formed during thermal cycling.[4] Examples are the use of radioactive tracers, such as the uptake of $^{32}P$ or $^{14}C$ as a function of the intensity of thermal cycling. Another proposed method involves the detection by luminometry of ATP synthesized during thermal cycling. Michael Kaufmann in a first experimental search for thermosynthesis, has tried to detect ATP synthesized after thermal cycling of a candidate thermosynthesizing protein from *Aquifex aeolicus*, probably one of the oldest bacteria.[5] These direct methods are called bottom-up approaches, since one works with enzymes, which are more simple than organisms. Already known and identified enzymes may be investigated

A top-down demonstration would, in a first step, involve showing a benefit of thermal cycling for whole organisms. The second step would constitute the unraveling of the mechanism of the benefit, in the hope of identifying a partial process based on energy gain from thermal cycling. Such an approach would resemble the standard working method in biochemistry, in which one takes a physiological function such as photosynthesis and respiration, and then works one way down to find the partial molecular processes. In the case of the just mentioned physiological functions, finding these partial processes has taken many decennia.

Compare thermosynthesis to nitrogen fixation. In plants nitrogen fixation is a complex process that requires symbiotic anaerobic bacteria; the anaerobiosis may be a relic of reducing conditions on the early Earth. Similarly the numerous thermoperiodic phenomena in biology may be a relic of the origin of life. After the transition from living on thermal cycling to living at constant temperatures on light or food, some enzymes in today's organisms may have kept the ability to make use of thermocycling. When the temperature remains constant, the organism would have to spend much energy, using complex physiologies, to mimic thermal cycling. Applying thermal cycling to today's organisms then may be beneficial, since it would enable performing some key processes at a smaller energetic cost.  Most interestingly, a similar notion has already been proposed:  according to the 'McLaren hypothesis' the organisms that perform daily vertical migrations in natural, thermally stratified waters do so because they gain an energetic advantage from the associated thermal cycling.[6]

---

[1] Muller and Schulze-Makuch, Origins of Life and Evolution of Biospheres, (2006) in press.
[2] Muller, BioSystems, 82 (2005) 93.
[3] Muller, www.arxiv.org/physics/0501050
[4] Muller, Astrobiology 3 (2003) 555.
[5] Kaufmann, Klinger and Muller, Abstract 101 B, Astrobiology 4 (2004) 264.

    Microbiological methods especially are considered promising for the top-down method: a demonstration seems feasible of enhanced starvation survival of bacteria while they are thermocycled.  Again, such an effect would of course only be consistent with thermosynthesis, and proof would require the mentioned much more detailed investigation.  And even if thermosynthesis does not occur in bacteria, a study of the effect of rapid thermal cycling would be of interest on its own, and would be a continuation of studies from other authors on the effect of temperature on bacteria: rapid thermal cycling is an untried, easily applied, new potential handle on bacterial physiology. The offer by Susan Childers in May to use the facilities of her microbiology lab at the Geology Department of the University of Idaho for microbiological experiments was therefore gladly accepted.

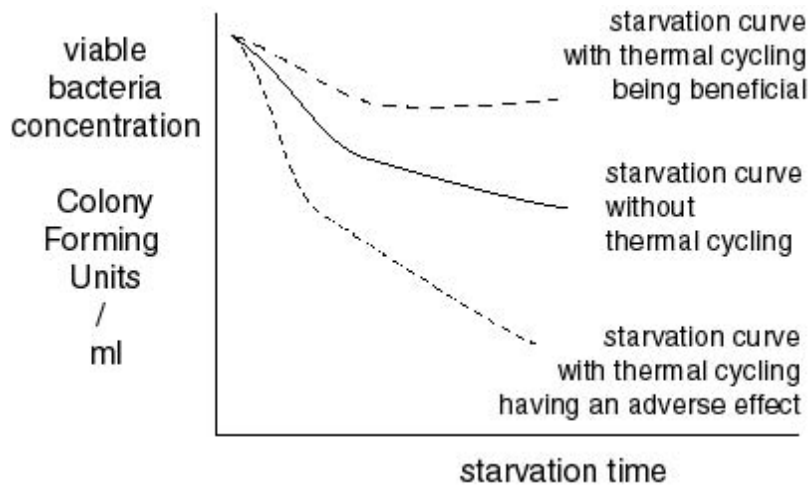

Figure 1 illustrates the proposed microbiological approach. Starvation is supposed to kill bacteria (or at least lessen their viability); their numbers should go down (although increases upon starvation have been reported as well).  Viable bacteria are counted by their Colony Forming Units (CFUs): bacteria are diluted until about 30 to 300 per 0.1 mL. This volume is put on an agar plate, which is incubated overnight in a 37 C oven.  The bacteria scattered over the agar plate multiply exponentially and overnight form visible and countable colonies.  The curve of CFU vs starvation time is called a starvation curve.  If thermal cycling were a stress factor, the starvation curve would go down with time due to the additional adverse factor; if it were beneficial, the curve would go up.

    In the use of starving bacteria the microbiological approach resembles the study of Gusev and Schulze-Makuch[7] on the growth of *Escherichia coli* in distilled water.  Therein the observed increase of CFUs of starving bacteria is linked to naturally occurring radio waves that have not been shielded by metal.  It was hoped to investigate this phenomenon as well, but lack of time as

---

[6] McLaren, American Naturalist 108 (1974) 91
[7] Gusev and Schulze-Makuch, Naturwissenschaften 92 (2005) 115

prevented this. The performed radio wave experiments are discussed in the **Radio Wave Experiments** section.

We describe hereafter both the starvation and radio wave experiments in more detail. Although the results are inconclusive, and considered insufficient for publication in a scientific journal, they are considered of sufficient interest to publish them on the Internet. They may easily be followed up. This document has therefore pedagogical purposes, and although it has the headings of a scientific article, the style is less concise.

The literature on starvation and the effects of thermal shock in bacteria is extensive. It is tempting to quote this literature extensively, but since investigating the effects of rapid thermal cycling is a new idea, and the existing literature is only partially applicable, we refrain from doing so. In a later stage the literature cannot be ignored, of course, but in this early, pioneering stage the emphasis is on finding new phenomena and reproducing them.

# MATERIALS AND METHODS

This section describes the constant part of the experiments; the variations in the method are topic of the RESULTS section.

*mediums used*
The composition of the mediums was taken from *Molecular Cloning*.[8]
The *LB medium* (Luria-Bertani Medium): 10 g/L tryptone, 5 g/L yeast extract, 10 g/L NaCl. pH was adjusted to 7.0 For pouring plates 15 g/L bacterial agar was added.
The *M9 minimal medium* consisted of 750 mL sterile $H_2O$ to which had been added: 200 mL 5x M9 salts solution (made in turn from 64 g/L $Na_2HPO_4.7H_2O$, 15 g/L $KH_2PO_4$, 2.5 g/L NaCl, 5.0 g/L $NH_4Cl$), 2 mL 1 M $MgSO_4$, 0.1 mL 1 M $CaCl_2$. According to protocol the components of the M9 medium have to be sterilized by autoclaving them separately. In addition some components were sterilized by filtration (0.22 μm filter): 0.1 mL/L chelated minerals, 1 mL/L Wolf's medium, and 0.1 mL/L 20% glucose.[9]
The *PBS medium* was used for dilution of bacteria. It contained 8 g/L NaCl, 0.2 g/L KCl, 1.44 g/L $Na_2HPO_4$ and 0.24 g/L $KH_2PO_4$.

Susan Childers recommended the use of the bacterium *E. Coli* strain K12, a standard bacterium in microbiological research, also used by Gusev and Schulze-Makuch in their study (ref. 7).

The working method was the following. A sample of the bacteria strain was taken from the storage container in the freezer, and streaked out and diluted on an agar plate in the standard manner. The plate was put in a 37 C oven overnight, whereafter it was put in the fridge. A good looking (= good circular symmetric) colony from this plate was taken for bacterial growth during the following months.

When growing a new batch of bacteria, the colony was put in a tube with a few mL (3 – 10) mL of LB medium and the tube was put overnight in a shaking bath at 37 C. When bacterial growth was visible, 0.1 mL of the LB medium was added to an erlenmeyer containing 100 mL of glucose-limited M9 growth medium. After a few hours the turbidity indicated bacterial growth and the $OD_{666}$ (optical absorption at 666 nm) was tracked. When the bacteria had consumed all glucose, growth ended; this occurred typically in the $OD_{666}$ range 0.1 – 0.2. The day of the onset of stationarity was counted as 'day 0.' The next morning the medium with the stationary bacteria was either directly diluted in the starvation medium (Runs 1, 2 and 3), or the bacteria therein were first repeatedly washed by the starvation medium, and then diluted (Runs 4, 5, 6 and 7). The reason for dilution (by a factor 100) was to lessen the chance of the occurrence of cannibalism.

Typically 1 mL of the growth medium with the bacteria was added to the starvation medium, which had the same composition as the growth medium minus glucose, chelated minerals and Wolff's medium. Aliquots of 0.2 mL of the starvation medium were added to ~ 100 PCR tubes. By using many PCR tubes, an infection in one tube cannot infect another, and, more generally, tubes cannot affect each other.

The filled PCR tubes were exposed for several days to various thermal treatments. The cyclic temperature (CT) was effected by a PCR thermocycler. Constant temperatures were applied at 35 or 37 C (oven temperature OT), at room temperature at 20 C (RT20) or in the fridge (FT04). In Runs 5, 6 and 7 constant temperatures of 24, 27 or 37 C were effected by heat baths.

---

[8] Sambrook and Russell, *Molecular Cloning. A Laboratory Manual*, Appendix 2, section A2.2
[9] Schult et al, Journal of Bacteriology 170 (1988) 3903. In this study carbon starvation was effected by a glucose concentration of 0.025 % glucose wt /vol = 25 mg / 100 mL, which resulted in an $OD_{666}$ of ~ 0.3, equivalent to 3 x $10^8$ cells/mL.

After the treatment, 0.1 mL of the 0.2 mL in the PCR tube was taken, diluted and plated. For instance 0.1 ml was added to a 9.9 ml dilution tube (dilution by 100), from which 1 ml was added to a 9 ml tube (dilution by 10). When from the last tube 0.1. ml was taken, the conversion factor from plate to CFU/ml in the original solution equals $100 \times 10 \times 10 = 10^4$. Using the dilution data, the plate counts were converted to CFU or viable cells/mL, and plotted against exposure time (note that the time range on these plots varies).

In our pioneering study, we started with a search for an effect, with the intention to increase the accuracy and precision after an effect had found; unfortunately that stage was not reached. For optimal precision it is commonly advised to ignore plate counts outside the range 30 - 300. Since however differences by a factor 2 in plate counts in otherwise identical conditions were regularly found, even when counts did fall in the range 30 - 300, and plate counts cannot always be predicted, we used data below 30 or above 300, as long as counting was possible. We are aware that this constitutes an inaccuracy.

During the analysis of the data a few outliers were met (in R2.2 an outlier is shown) and removed. The identification of a data point as outlier is somewhat subjective, and in a follow up study a defined procedure for removing them should be followed.

Halfway the studies it was found that the dilution tubes could incur a —highly variable but significant— weight loss during sterilization (up to 5%). The determined plate counts were corrected for this, using an Excel spreadsheet and the weights of the liquids in the dilution tubes recorded after dilution.

*Radio wave experiments*
The radio wave setup consisted of a 6060A Fluke Synthesized RF Signal Generator which can produce the high frequency power. The frequency range is 100 kHz to 1050 MHz. Two antennas permit the sending and receiving of radio waves. A HP Model 478A Thermistor Mount connected to a HP 432A Power Meter was used to measure the received power.

# RESULTS

**Run 1** (June 20 – July 6)

In this run the CT treatment consisted of 10 min 37 C: 10 min 20 C. The PCR machine often crashed, making frequent restarts necessary, which obviously was not always possible, for instance not at night. Not much value should therefore be given to the CT results of this run.

At the time it was considered unimportant, and therefore not recorded, whether the components of the growth medium were sterilized separately or not.

The starvation medium did not contain $Ca^{2+}$.

The table gives the CFUs (in thousands / mL) calculated from the plate counts:

| Day | CT37:20 | OT37 | RT20 |
|---|---|---|---|
| 1 | | 2570 | |
| 2 | | | 320 |
| 3 | 520 | 630 | 600 |
| 4 | 440 | 390 | 560 |
| 5 | | | 800 |
| 6 | 19 | 10 | 33 |
| 8 | 11 | 16 | 13 |
| 14 | 150 | 14 | 279 |
| 30 | | | 28 |
| 42 | | | 1250 |
| 43 | | | 560, 1230, 2030, 2210 |

Viable bacteria concentration in thousands of CFU/mL

The same data is shown in the following plots. The results of this first run are given mainly for completeness; the plots from the following runs give much more information.

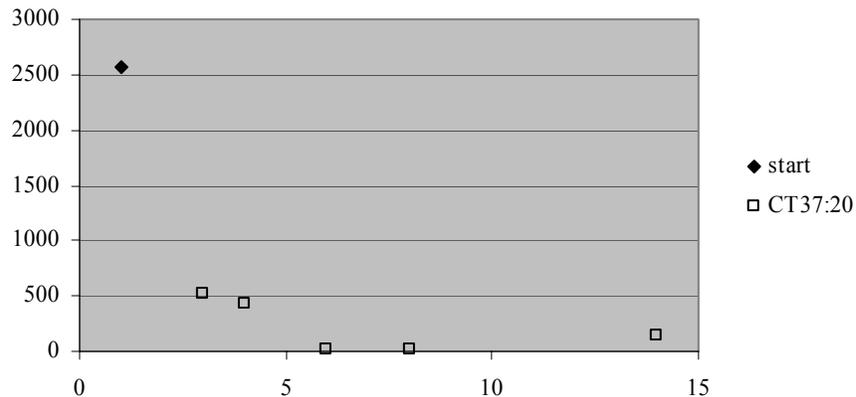

Fig. R1.1: Application of thermal cycling (37:20) treatment.

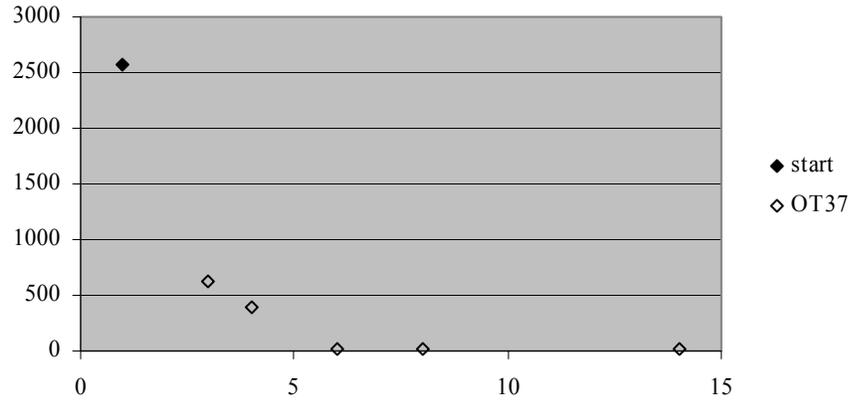

Fig. R1.2: Application of constant oven temperature (37 C) treatment.

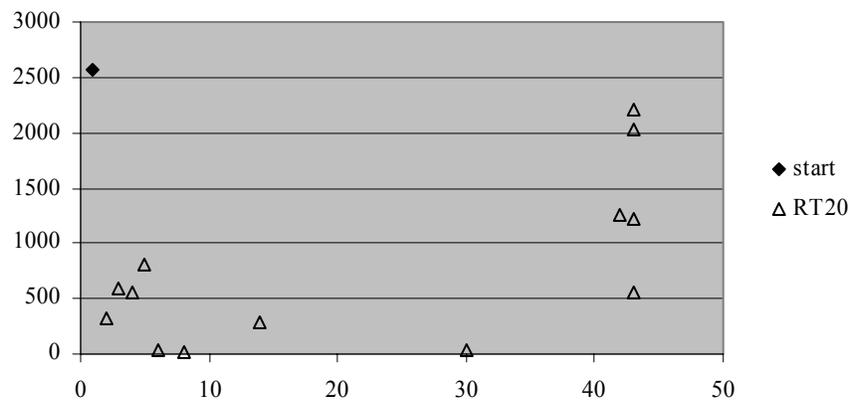

Fig. R1.3: Application of constant room temperature (20 C) treatment.

The RT20 treatment shows a return of viability after 6 weeks. This phenomenon was not further investigated. The focus in this study is on the events during the first one or two weeks.

The combined results of the three treatments during the first week is shown on the following plot:

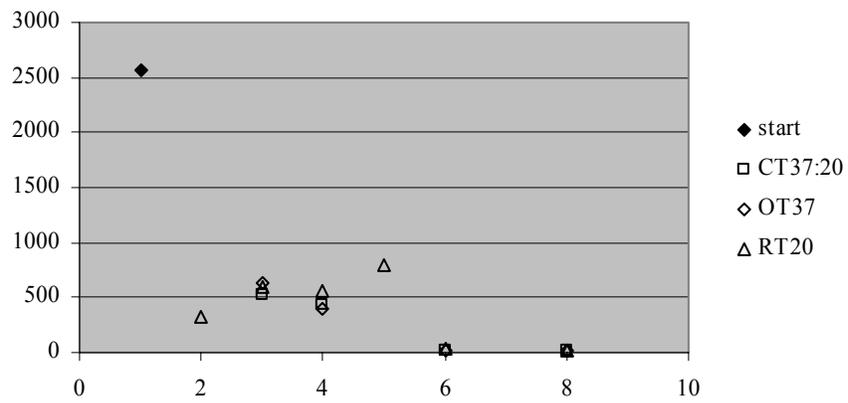

Fig. R1.4: Combined results of the various treatments on days 1 to 8.

At days 2 to 6, the time of platings of different treatments, the CFUs were identical. All treatments result in a CFU decrease. The data from this run do not yield evidence that the CT treatment increases the CFU compared to OT or RT treatments.

**Run 2** (July 23 – August 6)

The PCR machine was repaired.

The thermal cycle was halved to 5 min: 5 min, as shortening the cycle time was expected to increase the chance of an observable effect. The PCR machine can be programmed to a maximum of 99 cycles; after starting up, the machine runs for 18 – 20 h. Upon a start late in the afternoon, it has to be restarted again the following morning in order to effect continuous cycling. Restarting was not always feasible, so on some days there was no thermal cycling for a few hours and the tubes then remained at constant room temperature for a few hours.
The PCR machine cools by air. Cooling takes much time if the target cooling temperature is near room temperature. In order to shorten the thermal cycling time the range of the temperature cycle was shifted to above room temperature, to 37:27 C.

$Ca^{2+}$ was absent in the starvation medium.

As simulation of the Gusev and Schulze-Makuch experiments with radio waves (Ref. 7) some of the tubes kept at RT were wrapped in aluminum, the metal supposedly screening the tubes from radio waves by acting as a kind of Faraday cage (RT20A treatment).

| Day | CT37:27 | OT37 | RT20 | RT20A (aluminum foil) | FT04 |
|---|---|---|---|---|---|
| 1 | 1530, 2010 | | | | |
| 2 | 1940 | 2110 | 750 | 420 | 1010 |
| 3 | 1560 | 1400 | 360 | 720 | 790 |
| 4 | 1160 | 600 | 198 | 229 | 221 |
| 5 | 930 | 470 | 294 | 830 | 266 |
| 6 | 1940 | 920 | 163 | 500 | 86 |
| 8 | 1930 | 5710 | 810 | 580 | 98 |
| 11 | 1000 | 480 | 174 | 490 | 146 |
| 14 | 2340, 3130 | 1940, 2560 | 268, 570 | 557, 1220 | 134 |

Viable bacteria concentration in thousands of CFU/mL

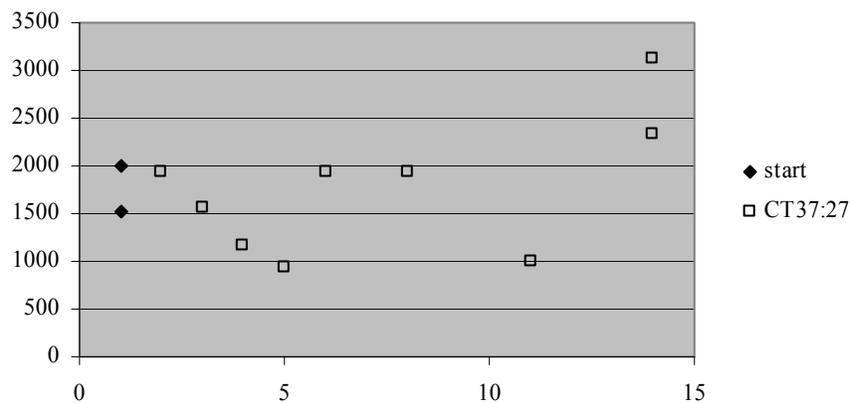

Fig. R2.1: Application of thermal cycling (37:27) treatment.

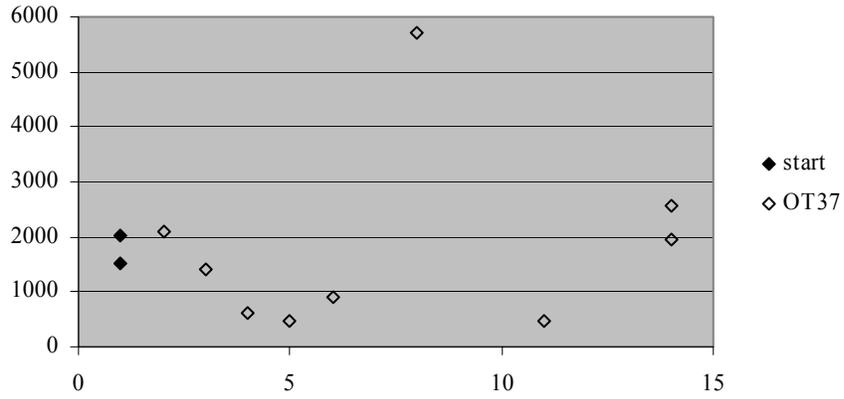

Fig. R2.2: Application of constant oven temperature (37 C) treatment.

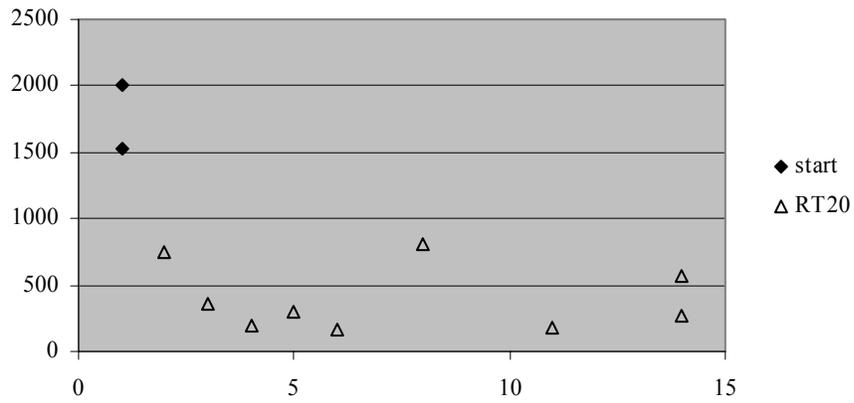

Fig. R2.3: Application of constant room temperature (20 C) treatment.

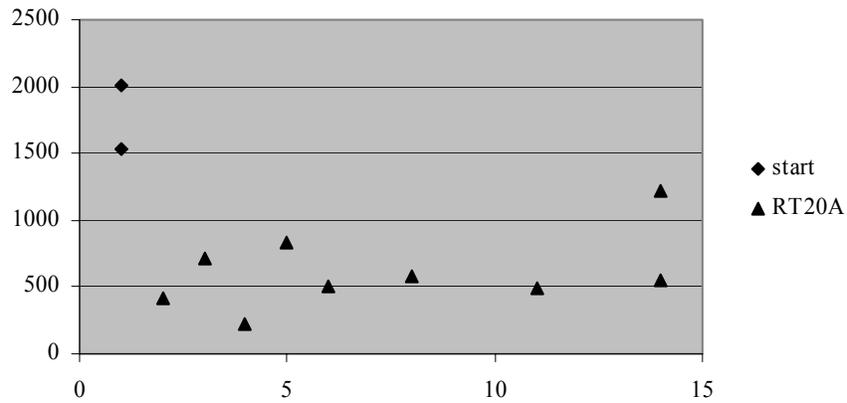

Fig. R2.4: Application of constant (room) temperature (20 C) treatment, with tubes wrapped in aluminum foil.

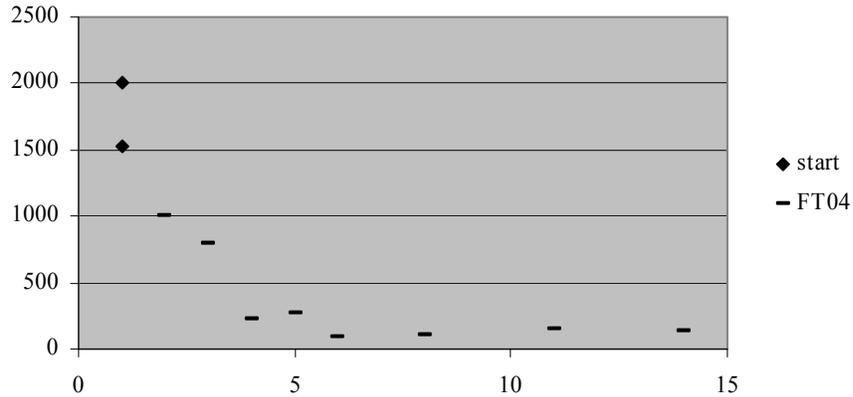

Fig. R2.5: Application of constant fridge temperature (4 C) treatment.

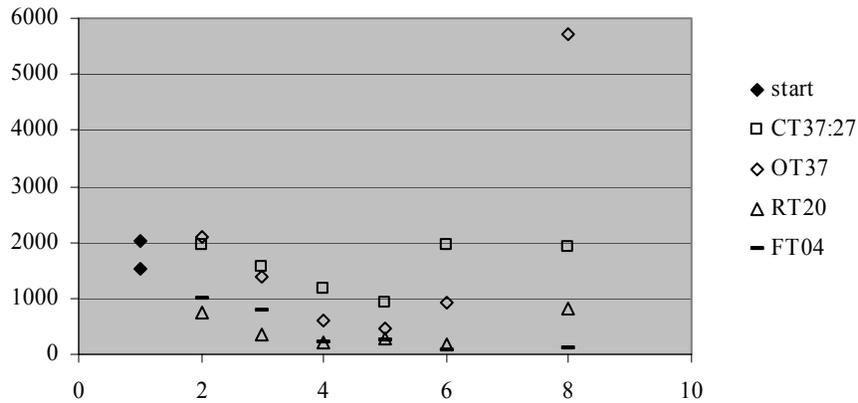

Fig. R2.6: Combined results of the various treatments on days 1 to 8.

The CT treated tubes (R2.1, R2.6) do not show a decrease in CFU. CT tubes show higher counts then RT and OT37 tubes, in agreement with the proposed beneficial effect of thermo cycling as shown in Figure 1.

The OT37 tubes (R2.2) show a decrease in CFU until day 6, when the counts stabilize. The FT04 tubes (R2.5, R2.6) show a decrease, a decrease that at day 8 is larger than the decreases of the RT20 tube, an unexpected result. The counts of RT20 tubes in turn are consistently lower than those of the OT37 tubes.

The RT20A treatment seems to result in higher counts (about 500) than the RT20 treatment (under 500), which is in contrast with the Gusev – Schulze-Makuch findings (Ref. 7) that screening off of radio waves leads to a decrease in counts. .

There is an overall increase in the CT and OT37 values in the second week.

These results are considered consistent with a beneficial effect of the CT treatment.

**Run 3** (August 18 - September 1)

After discussion with Dirk Schulze-Makuch and Susan Childers, the following changes were made:
1. Two plates were made every day; on the first, the sixth and the fourteenth day, several plates were made.
2. One cause of variation may be a variable number of bacteria put in the tubes during filling, as bacteria may sink to the bottom of the erlenmeyer containing the starvation medium. For this reason the erlenmeyer was better stirred. During pipeting liquid was drawn up and let fall back three times to improve mixing as well.
3. In order to avoid the condensation that occurred in the plates during Run 1, the plates were kept in the oven all time during Run 2. This however dries them out, and may lessen their sensitivity as this dryness makes it more difficult for bacteria to grow. Plates were not kept in the oven any more, but were dried for about an hour in the oven before use.

In Runs 1 and 2 a shortcut had been applied to the preparation of the growth and starvation mediums: the components had been added together in a single solution, and had then been sterilized. As a result the medium acquired a milky look, which is attributed to a precipitate of $Ca^{2+}$ with phosphate and/or sulfate. Therefore these components were now sterilized separately first and mixed next, as prescribed by the protocol in Ref. 8, which states that the mixing has to occur below 50 C.

When a precipitate is *not* formed in a solution because of a low temperature, but the precipitate is formed at a higher temperature, this quenching of precipitation implies the low-temperature stability of a chemical disequilibrium, from which organisms could in principle gain free energy. Direct coupling of the disequilibrium to ATP synthesis can easily be imagined, for instance by using a $Ca^{2+}$-ATPase.

| Day | CT37:37 | OT37 | RT20 | FT04 |
|---|---|---|---|---|
| 1 | $98 \pm 47$ (n = 21) (s.d. / av. = 0.48) | | | |
| 2 | 516, 867 | 531, 677 | 405, 407 | 77, 188 |
| 3 | 360, 1160 | 570, 630 | 262, 291 | 110, 126 |
| 4 | 1270, 2270 | 131, 279 | 125, 224 | 100, 111 |
| 5 | 1590 | 151, 280 | 57, 283 | 73 |
| 6 | > 1000 (estimate) | $359 \pm 202$ (n = 15) (s.d. / av. = 0.56) | $164 \pm 87$ (n = 15) (s.d. / av. = 0.53) | $57 \pm 22$ (n = 10) (s.d. / av. = 0.39) |
| 7 | 227, 1750 | 420, 760 | 302, 561 | 13, 93 |
| 9 | 1100, 1170 | 340, 920 | 231, 410 | 61, 104 |
| 12 | 5500 | 350, 1280 | 1010, 1240 | 135, 158 |
| 14 | $3600 \pm 1000$ (n =4) (s.d. / av. = 0.28) | $960 \pm 560$ (n = 11) (s.d. / av. = 0.58) | $400 \pm 180$ (n =10) (s.d. / av. = 0.45) | $44 \pm 31$(n = 8) (s.d. / av. = 0.70) |

Viable bacteria concentration in thousands of CFU/mL

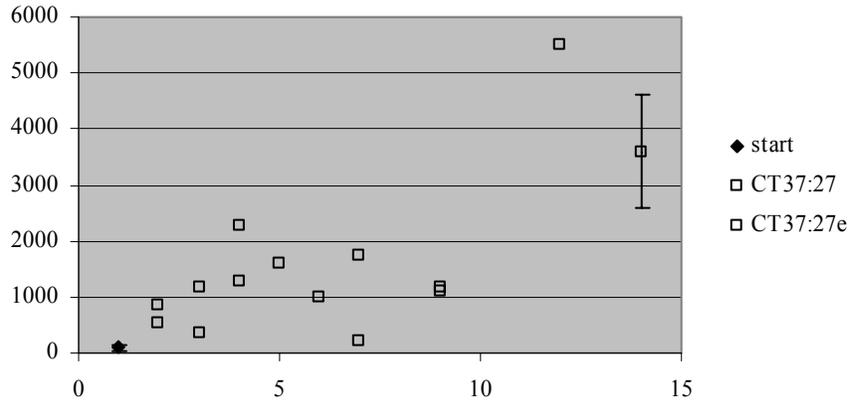

Fig. R3.1: Application of thermal cycling (37:27) treatment.

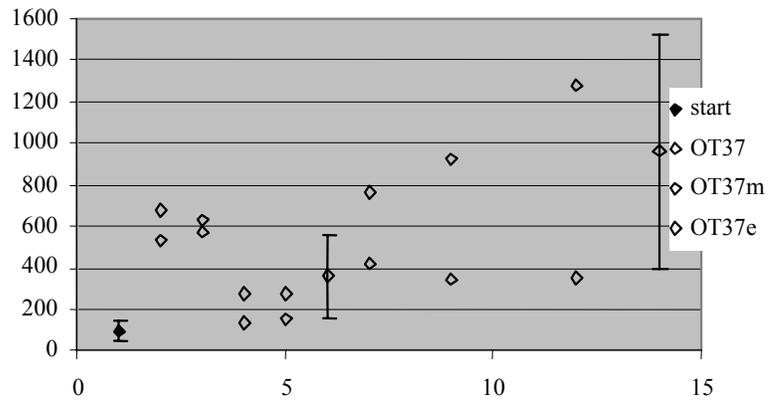

Fig. R3.2: Application of constant oven temperature (37 C) treatment.

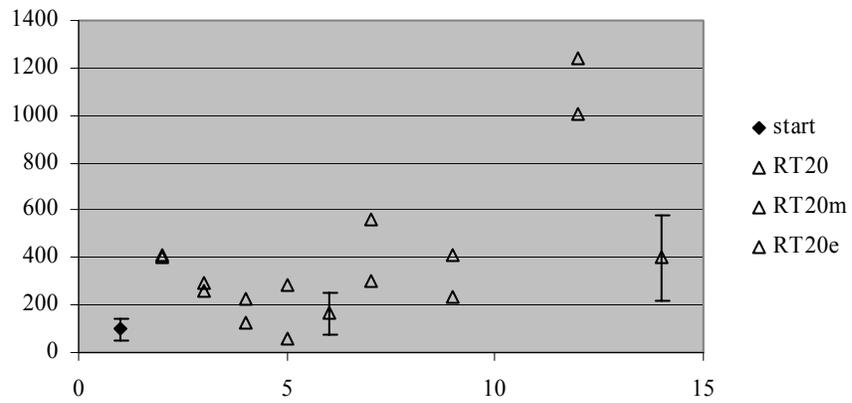

Fig. R3.3: Application of constant room temperature (20 C) treatment.

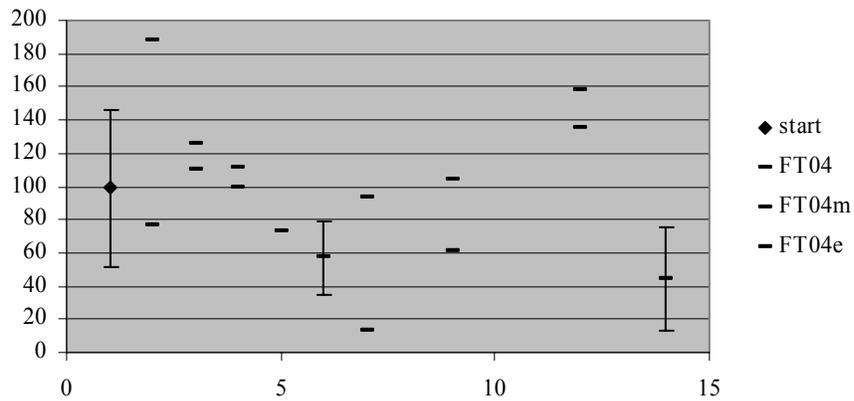

Fig. R3.4: Application of constant fridge temperature (4 C) treatment.

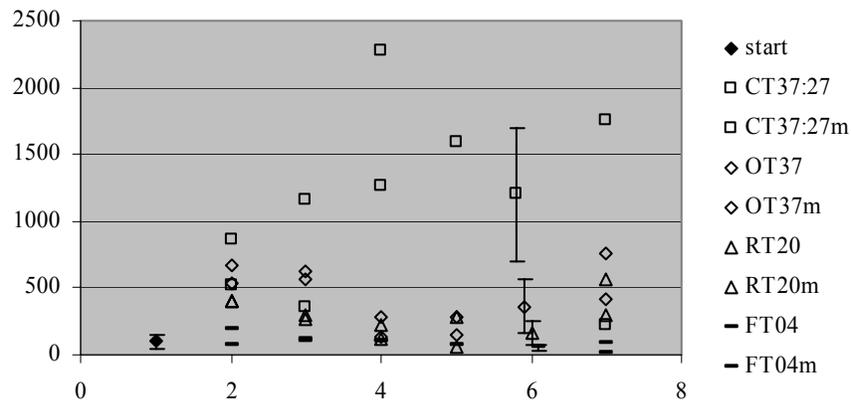

Fig. R3.5: Combined results of the various treatments on days 1 to 8.

On the 6th day many platings were made for all treatments. Due to oversight no additional dilution was done for the CT plates. Most of the counts were estimated to be far above 1000. The few counts below 1000 were interpreted as the low end tail of an average above 1000. The counts at day 6 in R3.1 and R3.5 are therefore estimates.

The key issue is the large variation in the counts. A main parameter is the ratio of standard variation and the average, which is about 0.5 for the tubes on day 1, which have not undergone any treatment. The platings at days 6 and 14 show similar values. One expects this ratio to go up following the treatments, and not to go down. The scattering is much wider than the Poissonian (statistical) variation of $\sqrt{n}$ that can be expected for plating. For instance, for a CFU of 25 the Poissonian variation is 5, for a CFU of 400 it is 20. Regularly, however, counts between plates differ by a factor 2. As long as the cause of this large variation is unknown one should be very careful to draw conclusions from effects of changes in experimental parameters, as these changes may affect this unknown cause instead of the proposed energy generator.

In contrast to Runs 1 and 2, the average counts are increasing for all treatments, except for the FT04 treatment.

This increase resembles the results of Gusev and Schulze-Makuch. Note however that in their study the bacteria were kept in distilled water, whereas here the bacteria are kept in a buffer.

The CT treatment shows the largest increase, at the 14$^{th}$ day the counts have increased by a factor 40. The OT37 treated tubes show an increase by a factor 10 after 14 days. The RT20 treatment results in an increase by a factor 4. The FT04 treatment shows an initial increase, but at days 6 and 14 the CFU have gone down.

In this run the CT treatment indeed resulted in a higher count.

**Run 4** ( September 10 - September 19)

This run was broken of after infection was detected.

**Run 5** (September 20 – September 28)

In this run the bacteria are washed after having reached the starvation stage; washing involves centrifuging for 15 min at 3500 rpm, discarding of the supernatant and resuspension of the pellet containing the bacteria in 20 ml of the starvation medium. The washing is performed three times, after which the bacteria are put in the starvation medium from which the PCR tubes are filled.

The dilution procedure consists of taking a sample from the PCR tube, and putting it in the first dilution tube. From this a sample is transferred to the second dilution tube, from which in turn the plated sample is taken. In this run the volume transferred between the dilution tubes was varied in such a way that the plate count was between 30 and 300.

The weights of the liquid in the dilution tubes were recorded after plating.

| Day | CT37:28 | OT37 | OT28 | RT20 |
|---|---|---|---|---|
| 1 | 525 ± 70 (n = 7) (s.d. / av. = 0.13) | | | |
| 2 | 1148 | 1072 | 546 | 206 |
|   | 1425 | 1772 | 566 | 387 |
| 3 | 985 | 873 | 638 | 248 |
|   | 2758 | 1077 | 701 | 498 |
| 4 | 1193 | 371 | 345 | 158 |
|   | 2613 | 971 | 533 | 1013 |
| 5 | 1149 | 133 | 452 | 149 |
|   | 3377 | 176 | 486 | 418 |
| 6 | 436 | 29 | 241 | 164 |
|   | 1072 | 288 | 264 | 307 |
| 7 | 418 | 96 | 492 | 206 |
|   | 752 | 243 | 549 | 589 |
| 8 | 748 ± 453 (n = 7) (s.d. / av. = 0.61) | 137 ± 78 (n = 8) (s.d. / av. = 0.57) | 575 ± 275 (n = 7) (s.d. / av. = 0.48) | 362 ± 214 (n = 7) (s.d. / av. = 0.59) |

Viable bacteria concentration in thousands of CFU/mL

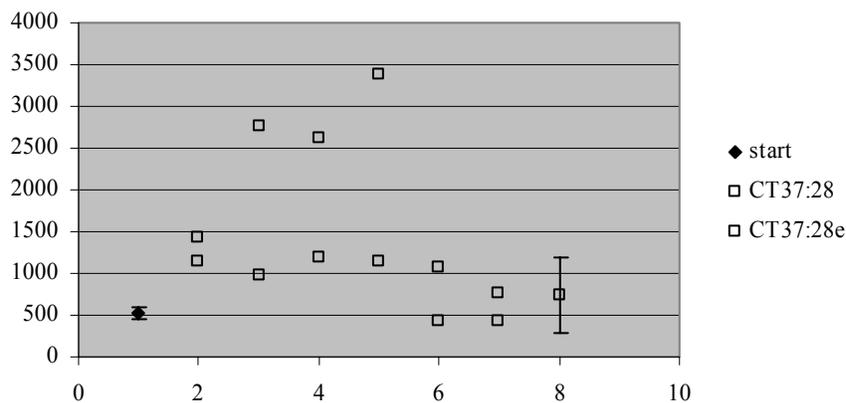

Fig. R5.1: Application of thermal cycling (37:28) treatment.

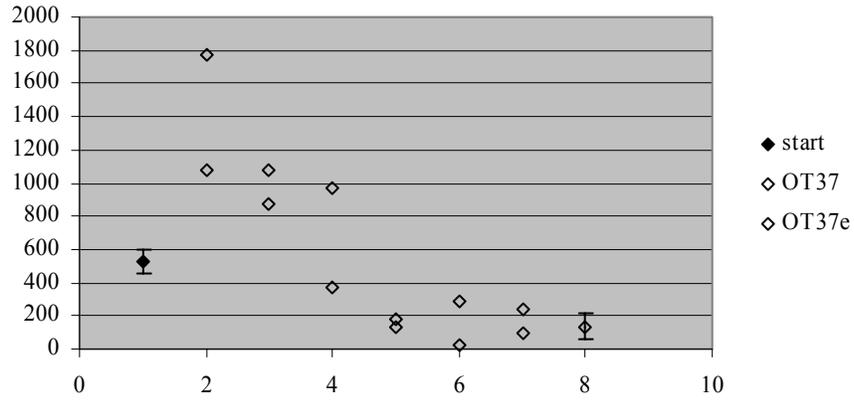

Fig. R5.2: Application of constant oven temperature (37 C) treatment.

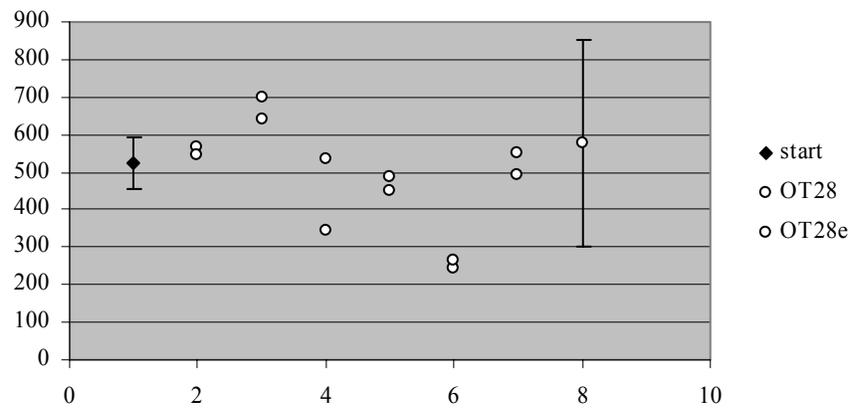

Fig. R5.3: Application of constant oven temperature (28 C) treatment.

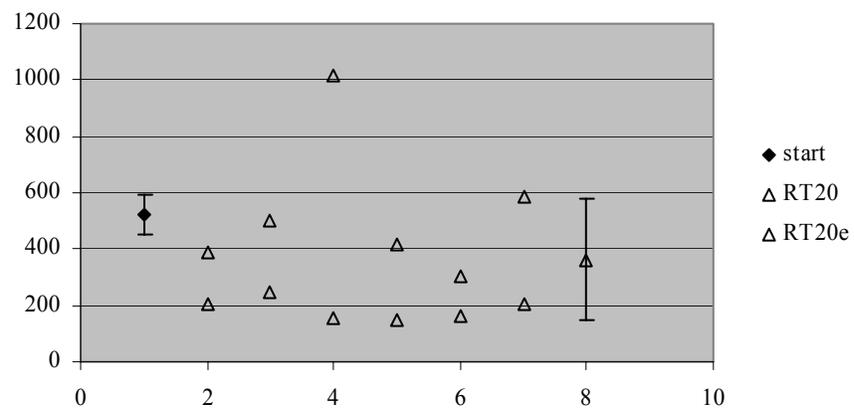

Fig. R5.4: Application of constant room temperature (20 C) treatment.

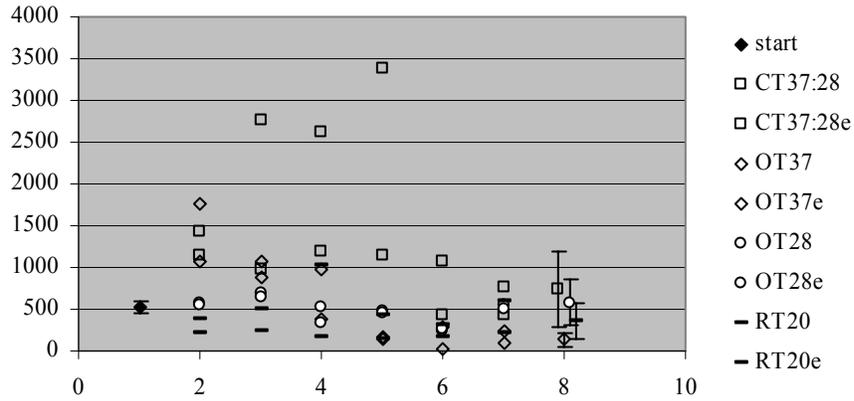

Fig. R5.5: Combined results of the various treatments on days 1 to 8.

The CT treatment (R5.5) resulted in strongly increased counts in the range 2500-3500, i.e. in an increase by a factor 6 in some, but not all, tubes during the first five days. Thereafter the values returned to the original value on day 1. At day 8 there is little difference with OT37 and OT28.

On day 2 both the CT treatment and the OT37 treatment result in an increase.

The large increases in counts after one week seen in Run 3 are absent here. The applied washing procedure may have removed traces of the carbon source that had remained in Run 3.

The beneficial effect of thermal cycling is pronounced in a few, but not all, CT tubes during the days 2 to 5.

**Run 6** (September 24 - October 2: this run overlapped with Run 5)

In this run no change was made compared to Run 5, the purpose being to reproduce the results of Run 5.

| Day | CT37:28 | OT37 | OT28 | RT20 |
|---|---|---|---|---|
| 1 | 194 ± 105 (n = 10) (s.d. / av. = 0.54 ) | | | |
| 2 | 643 / 879 | 448 / 602 | 341 / 953 | 33 / 93 |
| 3 | 1267 / 2202 | 1066 / 1480 | 690 / 701 | 112 / 171 |
| 4 | 872 / 1712 | 178 / 1903 | 647 / 899 | 170 / 267 |
| 5 | 2351 / 2967 | 61 / 132 | 702 / 1528 | 72 / 387 |
| 6 | 1062 / 1753 | 45 / 250 | 595 / 694 | 23 / 117 |
| 7 | 1970 / 2017 | 58 / 150 | 547 / 1027 | 96 |
| 8 | 708 ± 402 (n = 12) (s.d. / av. = 0.57 ) | 76 ± 36 (n = 15) (s.d. / av. = 0.47 ) | 437 ± 320 (n = 320) (s.d. / av. = 0.73 ) | 142 ± 62 (n = 11) (s.d. / av. = 0.43 ) |

Viable bacteria concentration in thousands of CFU/mL

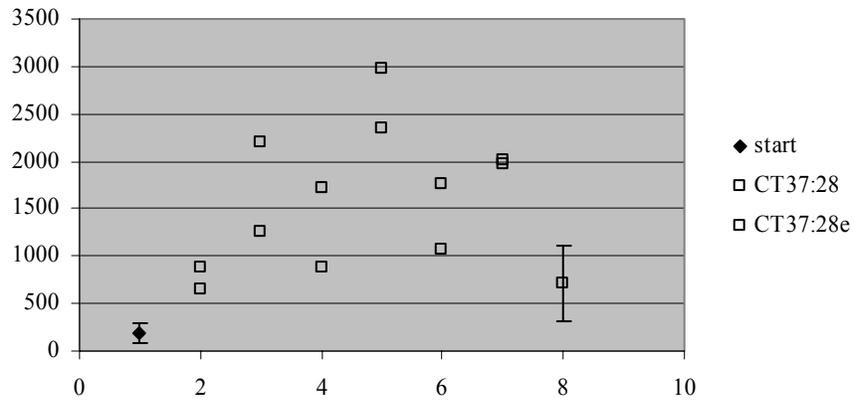

Fig. R6.1: Application of thermal cycling (37:28) treatment.

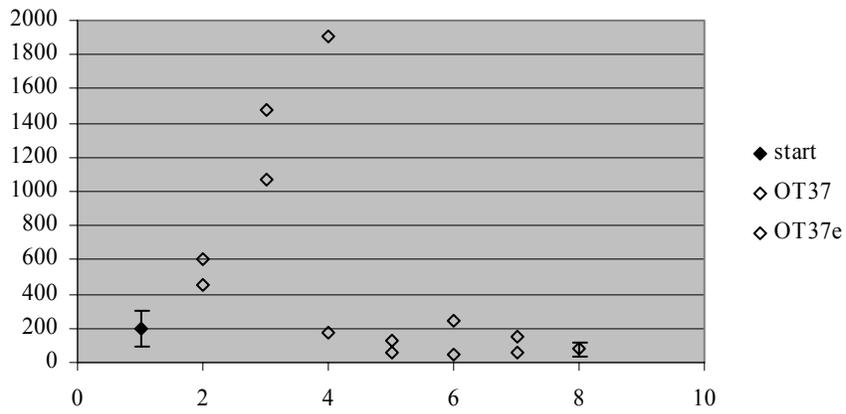

Fig. R6.2: Application of constant oven temperature (37 C) treatment.

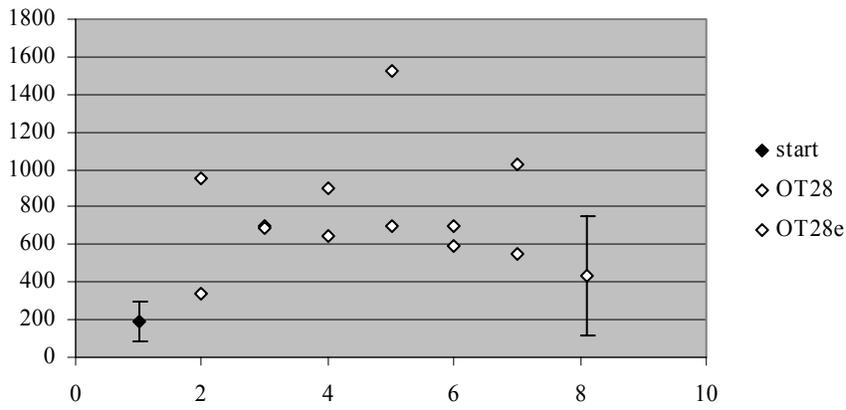

Fig. R6.3: Application of constant oven temperature (28 C) treatment.

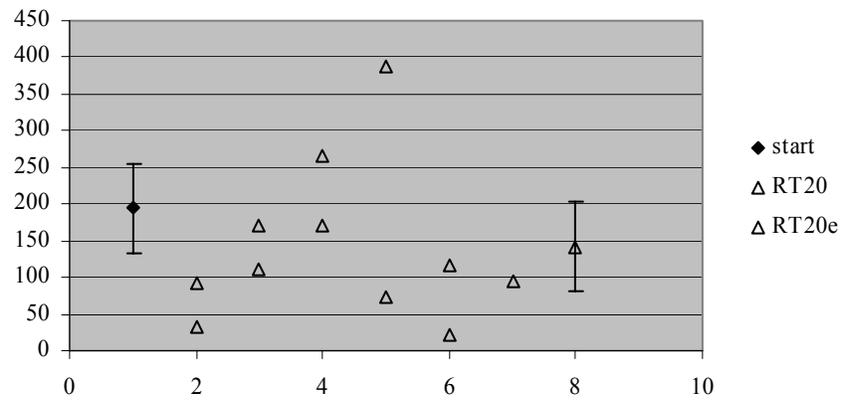

Fig. R6.4: Application of constant room temperature (20 C) treatment.

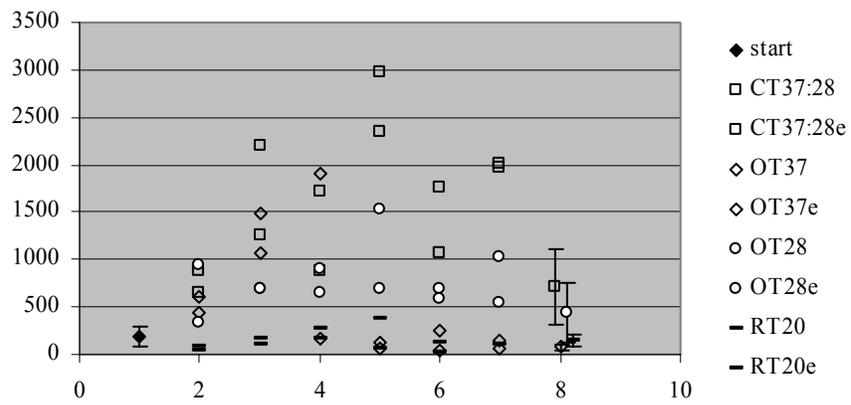

Fig. R6.5: Combined results of the various treatments on days 1 to 8.

Just as in the previous run there is an increase in CFUs until the fifth day for the CT treated tubes; there is however an increase for the OT37 treatments and OT28 until that day as well. The results are in good agreement with the previous run.

**Run 7** (November 4 – November 12)

At a meeting with Dirk Schulze-Makuch and Susan Childers it was decided to do another run, with modifications. The cycle range was increased to between 37 and 24 (extending the range to a lower value turned out to be impractical). Only two OT treatments were performed, at the range ends: 24 and 37 C.

A new method was applied to insert and remove the PCR tubes from the thermocycler. As a result the tube may have made a better thermal contact with the PCR machine.

| Day | CT37:24 | OT37 | OT24 |
|---|---|---|---|
| 1 | | 48 ± 25 (n = 10) (s.d. / av. = 0.52 ) | |
| 2 | 171 | 95 | 274 |
| | 238 | 234 | 319 |
| 3 | 682 | 853 | 561 |
| | 711 | 938 | 814 |
| 4 | 874 | 260 | 614 |
| | 1308 | 467 | 993 |
| 5 | 888 | 220 | 933 |
| | 947 | 312 | 949 |
| 6 | 513 | 341 | 718 |
| | 566 | 745 | 817 |
| 7 | 491 | 202 | 854 |
| | 539 | 323 | 1053 |
| 8 | 179 ± 58 (n = 5) (s.d. / av. = 0.32 ) | 200 ± 64 (n = 10) (s.d. / av. = 0.32 ) | 744 ± 140 (n = 9) (s.d. / av. = 0.19 ) |

Viable bacteria concentration in thousands of CFU/mL

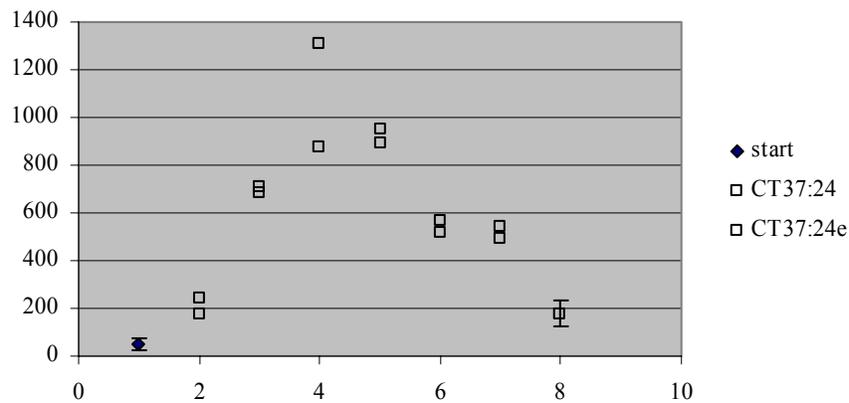

Fig. R7.1: Application of thermal cycling (37:24) treatment.

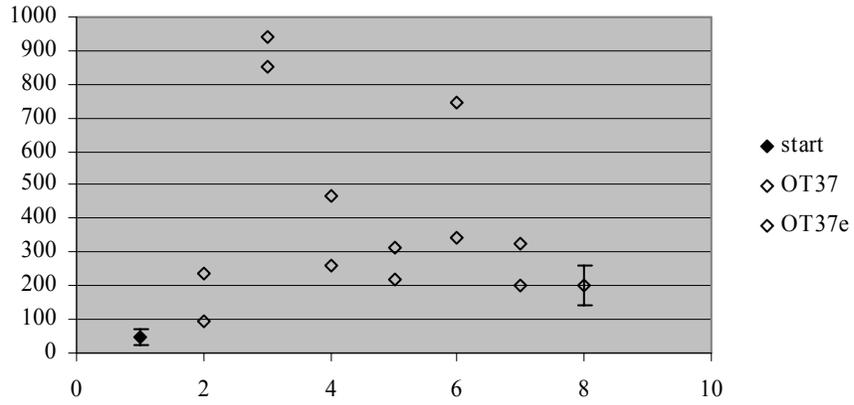

Fig. R7.2: Application of constant oven temperature (37 C) treatment.

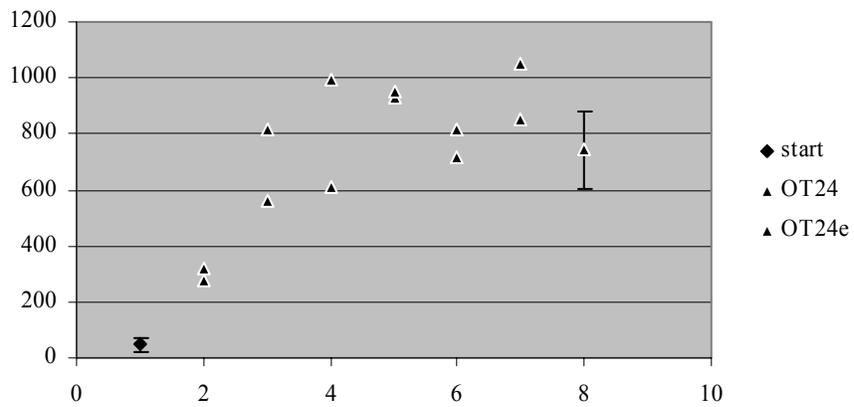

Fig. R7.3: Application of constant room temperature (24 C) treatment.

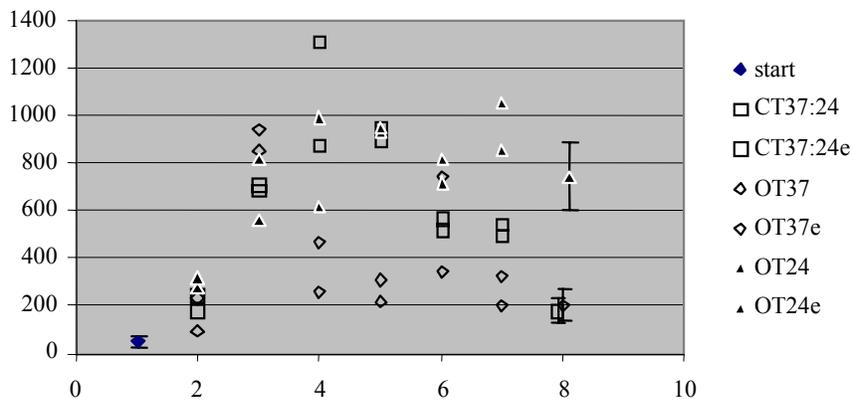

Fig. R7.4: Combined results of the various treatments on days 1 to 8.

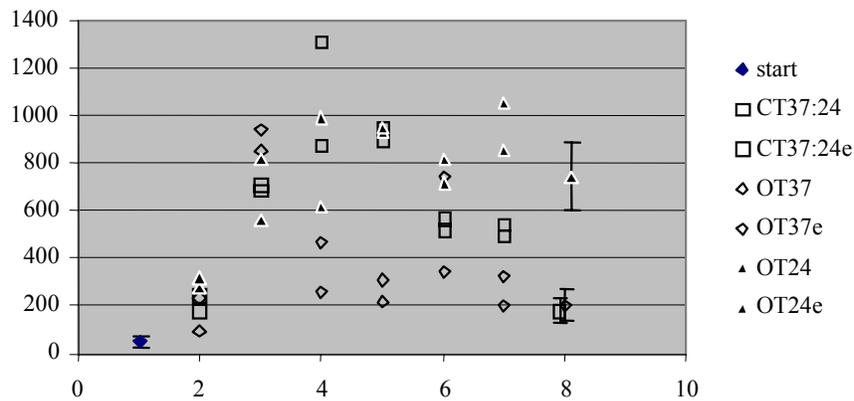

Fig. R7.5: Like R7.4, but with hand-drawn curves added.

Figure R7.1 shows little difference between the values of the duplicates, possibly because of the better thermal contact. The standard variation / average ratios are relatively low on day 8 as well, but on the first day the ratio still is about 0.5.
On day 4 the CT treatment has resulted in a 20 fold increase of the CFU, but then the increase collapses.

The expectation hat been that the OT24 treatment would yield values between those of the OT28 and OT20 treatments; they were however much higher, an unexpected result. This result suggests that it may be necessary to record starvation curves at more constant temperatures than has been done until now, in order to establish that a CT treatment gives different results than a properly time-averaged treatment over the temperature range of the cycle.

Two to five days after the onset of stationarity, thermal cycling results in an increase of CFUs compared to those of constant temperatures. Thereafter the CFUs decrease again to their original values: see R5.1, R6.1 and R7.1. The OT37 treatment, as shown by R5.2, R6.2 and R7.2, leads to an increase as well, but seems to decrease earlier.

Thermal cycling is seen to result in an increase in CFU count, which however also happens for the OT37 treated tubes.

Halfway this run Susan Childers made chromatograms of the proteins in the CT treated bacteria and in the constant temperature treated bacteria. No differences were seen.

*Study of variability*
Susan Childers suggested to get more information on the cause of the variability by repeatedly taking —necessarily small— samples of 20 μL from the same PCR tube, and plating these (only one dilution is needed).  This was done for several tubes:

|     | CT37:24 | | | | OT37 | | OT24 | |
| --- | --- | --- | --- | --- | --- | --- | --- | --- |
| Day | tube a | b | c | d | e | f | g | h |
| 3 | 245 | 142 | 137 | 135 | 93 | 177 | 202 | 104 |
| 4 | 344 | 128 | 159 | 175 | 109 | 78 | 206 | 159 |
| 5 | 338 | 151 | 166 | 160 | 54 | 130 | 120 | 133 |
| 6 | 493 | 178 | 228 | 171 | 47 | 65 | 33 | 67 |
| 7 | 235 | 150 | 269 | 246 | 56 | 81 | 230 | 148 |
| 8 | 569 | 336 | 194 | - | 76 | 111 | 258 | 294 |

plate counts — not CFU/ml

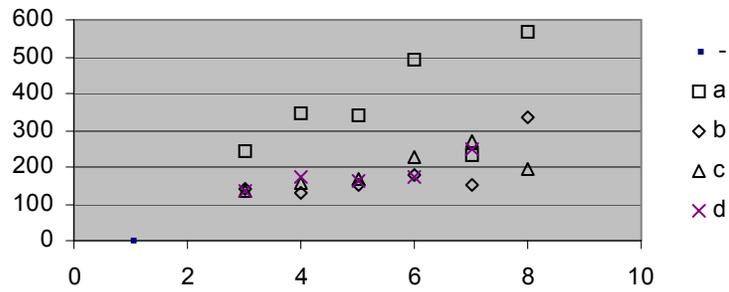

plate counts CT37:24 treatment

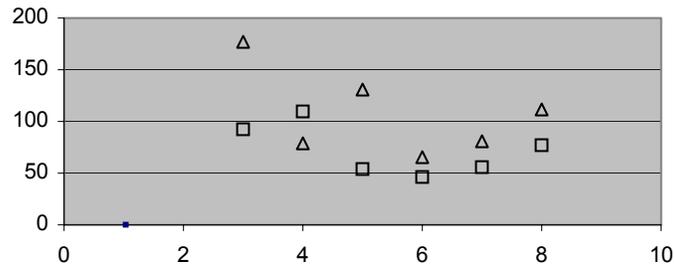

plate counts OT37 treatment

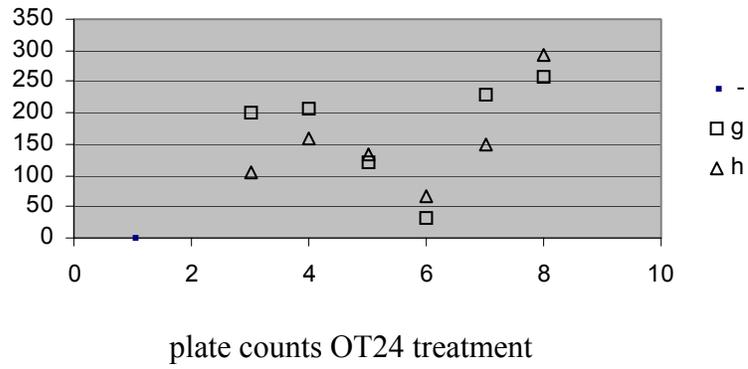

plate counts OT24 treatment

These plots show that a large part of the variation is intrinsic to the tubes.

Note that on the plot of the CT37:24 treated tubes in contrast to R7.1, no peak at days 3-4, followed by a collapse, can be discerned. The OT37 tubes show however a decrease with time similar to R7.2. The OT24 treated tubes show the same increase as R7.3

# Radio Wave Experiments

*The Gusev experiments*
We have already referred several times to the experiments by Gusev and Schulze-Makuch (ref. 7) on bacteria growth in highly-purified water under the condition that external radio waves can reach the bacteria (no shielding). In these experiments the same microorganism and M9 growth medium was used as in the present search for thermosynthesis. It was hoped to reproduce and to extend the Gusev experiments, but due to lack of time only some preliminary work was done.

It concerns two types of experiments. The first type started with a washed bacteria concentration of about $10^3$ CFU/ml in highly-purified distilled water. In the absence of an obvious chemical energy source, the concentration nevertheless reached the value of $10^5 - 10^6$ CFU/ml after two weeks. There was scattering in the growth: for instance, after one week in three of the samples shown the CFU concentration still was about $10^3$ CFU/ml, whereas in three other samples it was already about $10^6$ CFU/ml. This CFU increase could be impeded by storing the bacteria in a chamber made of permalloy[10]: after about 4 days the CFU concentration had gone down to nil. In these experiments the growth medium is in contact, by a set of filters, with open air.

The second type of experiment started with a high bacteria concentration, $\sim 5 \cdot 10^7$ CFU/ml, also in highly-purified water. The tubes were hermetically closed: there was *no* contact with the open air. The CFU concentration decreased linearly on a log (CFU) *vs* time plot, ending at $\sim 2 \cdot 10^5$ CFU/ml after 9 days. The permalloy had an adverse effect here as well. The CFY concentration went down faster when the tubes where kept inside the permalloy chamber (as $10^{-0.31\,d}$, equivalent to a half time of 1.0 days) than when they were kept outside the chamber in a thermostat (as $10^{-0.22\,d}$, $\sim$ half time 1.4 days).

Gusev and Schulze-Makuch give as explanation that the bacteria gain energy from external radio waves, and channel this energy into growth (the first type of experiments) or delayed death (second type). The permalloy would screen the external radio waves, and their beneficial effect.

What could be a mechanism for this beneficial effect? Biological energy conversion makes use of the chemiosmotic mechanism, in which for instance during photosynthesis or respiration a voltage is effected across a membrane. ATP is generated as protons move from the positive to the negative side of the membrane through the enzyme ATP synthase. It is proposed that radio waves transfer protons across the membrane against the potential, so that when the protons return ATP can be made by means of the chemiosmotic mechanism.

So the new question is, how could protons be pumped across the membrane? In distilled water the protons that result from the water dissociation reaction will function part of the time as free charges. In other media, for instance in fluorescent tubes, in the atmosphere at great heights, and in the Sun, such freely moving charges can behave as a plasma, a state of matter differing from the more familiar solid, liquid or gas states of matter. The ions in the plasma can undergo vibrations which are called 'Langmuir oscillations.' The angular frequency ω of the oscillations equals:

---

[10] Permalloy is a nickel iron alloy that is easily magnetized. It is therefore a suitable material for a Faraday cage, in which external electric and magnetic fields are screened off.

$$\omega = e \sqrt{(4 \pi n_H / m \varepsilon)},$$

where $e$ is the elementary charge, $n_H$ is the number of charge carriers per unit volume, $m$ is the charge carrier's (proton) mass and $\varepsilon$ is the dielectric constant.[11] Note that in the case of protons $n_H$ will vary with the pH. Reference 7 gives more detail of how, as the Langmuir oscillation is excited by the radio waves, the oscillating proton could gain enough energy to cross the membrane barrier of about 0.1 V. Making use of the relation $\omega = 2 \pi \nu$, where $\nu$ is the frequency, it follows that

$$\nu = e \sqrt{(n_H / \pi m \varepsilon)}.$$

At pH 7, $n_H = 10^{-pH} N_A$ (protons / L), where $N_A$ is Avogadro's number, equal to $6.022 \, 10^{23}$. Thus $n_H = 6.022 \, 10^{16}$ at this pH. Multiplication by the conversion factor 1000 L/m$^3$ then yields $6.022 \, 10^{19}$ protons /m$^3$. The frequency then equals $1.602 \, 10^{-19} \sqrt{(6.022 \, 10^{19} / (3.141 \times 1.673 \, 10^{-27} \times 80 \times 8.854 \, 10^{-12}))} = 0.644 \, 10^{9} \, s^{-1} = 644$ MHz.

At lower pH, say 6 or 5, this value must be multiplied with $\sqrt{10}$, respectively 10, yielding 2037, resp. 6440 MHz[12].

*Experiments at geology*
Somewhat arbitrarily, the just mentioned value of 644 MHz was chosen as frequency for the irradiation experiments. Note that the value in the signal generator used can easily be set to another value. When a pH measurement would yields 5.8, the proton concentration equals
$$10^{-5.8} = 1.58 \, 10^{-6} = 15.8 \, 10^{-7}.$$
Taking the square root of 15.8 yields 3.97; multiplication of 3.97 with 644 MHz yields 2557 MHz for the resonance frequency.

The frequency range of the generator is 100 kHz to 1050 MHz. The lowest pH of which the resonance frequency can be irradiated equals pH 6.6 (1050 / 644 = 1.63; squaring yields 2.6; multiplication by $10^{-7}$ yields $2.6 \, 10^{-7}$, which corresponds to a pH of 6.6). So at lower pH one may only be able to investigate the lower side band of the optimum resonance frequency. The intention was to radiate at say 644, 800, and 1000 MHz, and to see whether a found biological effect indeed increased when resonance frequency was approached.

The radio wave setup was tested by varying the frequency at constant emitted power of 19 dBm and recording the absorbed power indicated by the power meter.

---

[11] The values of $e$, $m$ and $\varepsilon$: the elementary charge $e$ equals $1.602 \, 10^{-19}$ C; the proton mass $m$ equals $1.673 \, 10^{-27}$ kg; the dielectric constant $\varepsilon = \varepsilon_r \varepsilon_0$, where $\varepsilon_r$ equals the relative dielectric constant, for water 80, and $\varepsilon_0$ equals the permittivity of the vacuum, equal to $8.854 \, 10^{-12}$ C$^2$ / N$^2$ m$^2$.

[12] This pH range of 5 – 7 follows from the statements that the pH would have to be larger than 5. Above pH 7 the significant presence of other positive ions follows from the need of having to compensate for the present OH$^-$ ion.

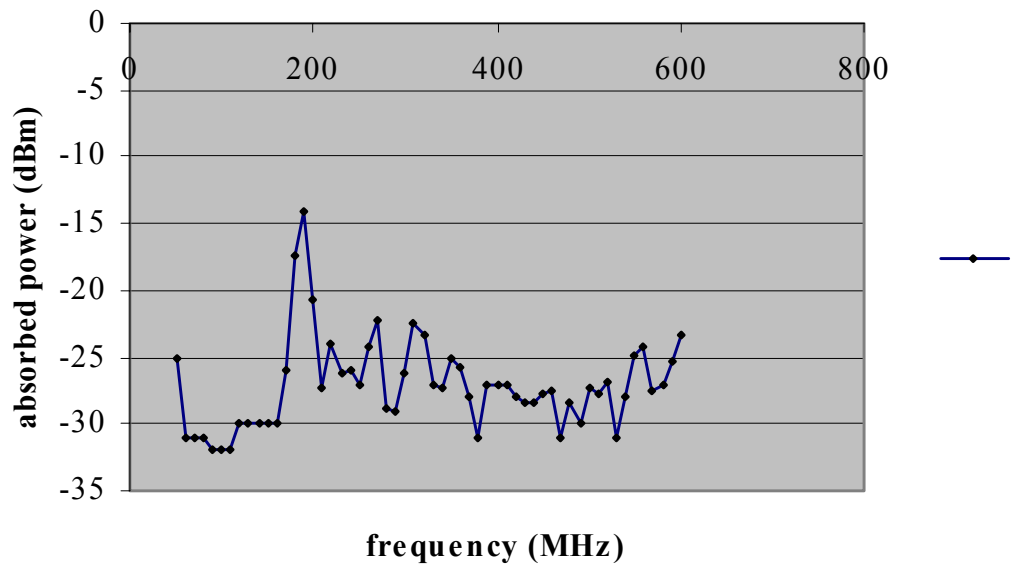

The figure shows the measured absorbed power in the absence of any absorbing substance between the antennas. Peaks are visible at ~ 50, 200, 270, 310, 350, 400 and 560 MHz. These peaks area associated with the shape of the antennas, which are of the not further explained 'log-periodic array' type (see google).
     The power transferred from the sender to the receiver depends therefore strongly on the frequency, which should be taken into account when investigating frequency dependency of a biological effect from using this type of set up.

The preliminary studies done were:
(a) during Run 1: a CFU count in a few (4) 42-days old PCR tubes;
(b) during Run 3: two experiments with larger numbers (12 and 10) of tubes.
In Run 3 the counts of the thermosynthesis studies were rather high; there was growth, possibly because of traces of nutrients in the medium. In Run 4 the bacteria were therefore washed. It was noted that bacteria washed in distilled water had a different color from those washed in the M9 medium. Unfortunately, there was no time left to subject the bacteria prepared in the manner of Ref. 7 to the radio wave type experiments. Although there are no results, the methods used in (a) and (b) may be of interest.

In (a) 5 41-day old tubes were irradiated, and the control group of 4 PCR tubes was not irradiated. The counts of the irradiated group were 157, 213, 352, 354 and 414: average 298 ± 106; the control had counts 56, 123, 203 and 221: average 151 ± 77. The difference is 147, not much smaller than the sum of the standard deviations of the counts of the two groups.
Due to computer failure some data (radiation time) of this experiment was lost.

During Run 3 the treated tubes were irradiated at maximum power (19 dBm) for three days while place at a distance of 0.62 m from the antenna. The controls were kept in the same room, at the larger distance of 5.5 m from the antenna.

The plate count of the irradiated tubes was 95 ± 29 (n = 12; sd / av = 0.31. dilution factor $10^3$);
for the controls these numbers were 105 ± 44 (n = 13; sd / av 0.42) .

A week later, the experiment was repeated.  The corresponding numbers:
plate count irradiated tubes 43 ± 17  (n = 10; sd / av = 0.40);
for the controls these numbers were 42 ± 20 (n = 10; sd / av 0.49).

There is no significant difference between the treatment and the control group.

Unfortunately the results of the experiments form Run 3 have limited validity, as there was such a large increase in CFU during this run, which is attributed to remaining trace amounts of nutrients. The absence of this increase in later runs is attributed to the washing of the bacteria in these later runs.
Another difference between the experiments in Run 3 and Gusev experiments is that in the former the bacteria are placed in a buffer, whereas in the latter they are placed in highly purified distilled water.

Clearly more research is needed, and no conclusions can be drawn at this moment.

# DISCUSSION AND CONCLUSION

During the first few days, say until day 4-5, there are large increases in count for the CT treatment, and to some extent for the OT37 and OT28 treatment as well. Here a beneficial effect of thermal cycling seems to be present. After a few days the increase in CFU however collapses.

*Overview of the runs*
As few platings were done and the thermocycler did not work well, the results of Run 1 have little relevance.

In Run 2 more platings were done, but still too few. R2.1 shows that the CT treatment did not result in a decrease in CFUs, whereas at constant temperature R2.2, R2.3, R2.4 and R2.5 show a decrease in CFU during the first five days.

In Run 3 many more platings were made, both in the beginning, halfway and at the end of the experiment. At day 6 counts were seen to increase the most for the CT treatments (R3.1; although not a precise value can be given for the colonies since there were too many too count), with smaller increases for the OT37 (R3.2) and RT20 (R3.3) treatments. The CFUs of the tubes kept in the fridge decreased (R3.4). There was a general large scattering in the results during this run, the standard deviation being typically about 50% of the average. Several attempts were made to enhance the accuracy.

Run 4 failed due to infection.

In Runs 5, 6 and 7 it was found that after one week (day 8) the thermally cycled tubes showed an increase by a factor 1.4 (Run 5), 3.6 (Run 6) and 3.7 (Run 7). The constant temperature treatments gave as results, at 37 C: a decrease by a factor 0.26 (Run 5), 0.39 (Run 6) but an increase of 4.2 (Run 7); at 28 C: an increase 1.1 (Run 5), 2.2 (Run 6); at 24 C: an increase of 15.5 (Run 7), and, at 20 C: decreases of 0.69 (Run 5) and 0.73 (Run 6). See the table:

|       | CT  | Constant 37 | Constant 28 | Constant 24 | Constant 20 |
|-------|-----|-------------|-------------|-------------|-------------|
| Run 5 | 1.4 | 0.26        | 1.1         |             | 0.69        |
| Run 6 | 3.6 | 0.39        | 2.2         |             | 0.73        |
| Run 7 | 3.7 | 4.2         |             | 15.5        |             |

ratio of CFU at day 8 and CFU at day 1

Whereas in Run 5 and 6 the CT treatment clearly resulted in a large growth compared to all constant temperatures, in Run 7 the constant 37 C and constant 24 treatment resulted in higher values than the CT treatment; in the case of CT34 in much higher values.
There is no clear pattern on day 8.

*The variation issue*
Obviously the large scattering in the result should be lowered. Much attention was given to find the cause of this scattering. The PCR tubes were boiled in order to remove traces of nutrients, the plates were properly dried (so that they were not too dry but also not too moist), the solutions containing the bacteria were mixed thoroughly, the streaking out of the bacteria on the plates was done uniformly.

On the Internet I found a document by the Alken-Murray Corporation titled 'Plate Count Procedure', on the 'Quality Control Method' for the 'Count of Aerobic Micro-Organisms.' Some quotes taken from the 'Notes about this procedure':

> Sampling error usually occurs because of an unequal distribution of cells in the culture or dilution fluid; the goal is to distribute the cells evenly by thorough mixing and then to obtain a representative sample in the sterile pipettor tip. Technical error is most often due to some inaccuracy in preparing dilution blanks or in pipetting technique. Error can be minimized by precise measurements in preparing blanks and by accurate pipetting technique, but it is impossible to avoid all error. Serial dilution and plating is at best an estimate of the number of live organisms present in the sample. Ideally, one organism should form a single colony, but cells may stick together to form chains and clumps, and they too will form a colony.

and

> The 95% confidence limit for spread plates containing 15 to 300 colonies is ±12% to ±37%. In practice, even greater variation may be found especially among results obtained by different microbiologists.

Large scattering in plate counts is a known phenomenon.

One explanation for the different results between the runs is that for each a different colony is used, and the bacteria in the colonies have mutated during their growth and subsequent starvation on the plate.

Mutation may also occur while the bacteria were initially grown during a run, before the onset of starvation. After starvation the bacteria grow however hardly at all, and it is not plausible for some of them to have acquired by mutation the ability to digest the plastic of the PCR tube, and that this selected for growth.

The experiments of Gusev are puzzling because of the large increase in CFUs in the absence of a known energy source in the distilled water medium. The present study is equally puzzling, because of the large CFU increase the first few days after the onset of starvation, followed by a collapse.

Changes in CFU can be associated with (i) a change in viability in a large, constant population of bacteria and (ii) with a real change in numbers of bacteria. In order to make this distinction the total of viable and non-viable bacteria should be counted, as can be done under the microscope.

*Quorum sensing*
The phenomenon of 'quorum sensing' shows that cultures of bacteria can have collective attributes: bacteria do not always behave independently. As a result of different onsets of negative feed-backs processes, the culture in one PCR tube may function very different from another. For instance, suppose that a high bacterial density, which may be local, is required for the secretion of a suppressor of division. When cells sink to the bottom of PCR tube, the less well mixed culture may locally reach the critical threshold density sooner than the well mixed culture, resulting in an overall lower number of bacteria in that less-mixed tube. Quorum sensing thus constitutes a random factor that can explain scattering, but only in general terms.

The maximal CFUs found for the thermal cycled bacteria in runs 5, 6 and 7 is about 1 - 3 $10^6$ CFU/ml. The similarity is surprising, since the start values of the these runs are 525, 194 and 48 $10^3$ CFU/ml.

Interestingly, an identical end value of about $10^6$ CFU/ml has also been observed by Gusev in his experiments on bacteria growth in distilled water, starting with $10^2$, $10^4$, $10^5$ and $10^6$ CFU/ml (Gusev: personal communication).

*General comments*
The M9 medium, even without glucose, is at a chemical disequilibrium, as shown by the formation of a Ca-precipitate at high temperature. In principle starving organisms could generate free energy from this disequilibrium. Calcium precipitation is a very important process in the environment, and is known to be microbiologically enhanced.

There is an extensive literature on starvation in bacteria, and it is known that bacteria can undergo morphological changes as a reaction, forming for instance very small bacteria. The viability can vary strongly as well.[13] The just mentioned small bacteria may use proteins inside the bacteria during the onset of starvation as energy source.[14]

*Demonstrating thermosynthesis*
The target of the present study was to find an effect of thermocycling. The found large increase in CFUs during the first few days after the onset of starvation is such an effect. In a follow-up the reproduction of the effect should be a first goal. When the phenomenon is reproduced, the standard methods of biochemistry and molecular biology, such as the detection of newly emerged proteins and nucleic acids, could be applied. This it could be established whether the phenomenon is related to thermosynthesis.

Thermosynthesis is at present mainly a theory for the origin of life and the emergence of biological energy conversion. It has a large explanatory power. The magnitude of the effect makes it difficult to investigate it experimentally. Many phenomena in the life sciences are much easier investigated, with corresponding larger chance of success.

This study is not the first experimental attempt to demonstrate thermosynthesis: Michael Kaufmann (Ref. 5) has already tried to detect ATP synthesis by thermal cycling a candidate protein isolated from *Aquifex aeolicus*, one of the oldest bacteria. The present study points to another plausible experimental handle on thermosynthesis.

Since *Aquifex* has a smaller distance to the origin on the map of the Tree of Life, it seems a more promising candidate for thermosynthesis than *E. coli*. There are many organisms in which thermosynthesis may occur, and many environments seem suitable for thermosynthesizers, both terrestrial and extraterrestrial. The methodology described in this study may therefore find a more general use.

---

[13] Roszak and Colwell, Microbiological Reviews 51 (1987) 365-379.
[14] T. Nyström, BioEssays 25 (2003) 204-211.

## Conclusion

In a pioneering study the effect of rapid thermal cycling on the CFU count of starving *Escherichia coli* was investigated. There is a slow increase during the first four or five days, followed by a collapse during the next few days. The most pronounced beneficial effect of thermal cycling seems to occur at the fourth day after the onset of starvation.

More research is needed to improve the method. Especially a decrease in the standard deviation of the measured CFUs is needed: a randomizing factor is present that could not be identified.

Microbiologists should be aware that even in the absence of glucose, the $Ca^{2+}$ ions present in the M9 medium commonly used results in a chemical disequilibrium that constitutes a potential biological energy source. This point was unfortunately realized late in the study.

This study shows that thermosynthesis can be investigated experimentally in whole organisms.

## Acknowledgements

I thank Dirk Schulze-Makuch for the given opportunity to do an experimental search of thermosynthesis, a risky endeavor. Susan Childers is thanked for offering the use of her laboratory and for many helpful discussions and suggestions. Ken Gordon is thanked for assistance with the PCR machine and for helpful advice on working with radio waves. Dirk Schulze-Makuch, Susan Childers and Victor Gusev are thanked for reviewing this document.

Anthonie Muller

Pullman, WA
April 10, 2006